\documentstyle[11pt,epsfig]{article}
\newcommand{\ffig}[4]{\begin{figure}[hbt]\vfill\begin{center}
            \mbox{\epsfig{figure=#1,height=#2}}\caption{#3}\label{#4}
            \end{center}\vfill\end{figure}}
\title{\normalsize\bf Note on a 'hint' for an annual modulation 
signature of a 60 GeV WIMP}
\author{ \normalsize G. Gerbier$^a$, J. Mallet$^a$, L. Mosca$^a$, C. Tao$^b$}
\date{October 13th 1997}
\begin{document}
\maketitle

\begin{flushleft}
{\footnotesize\it $^a$ DAPNIA/SPP, C.E.A. Saclay, F-91191 Gif-sur-Yvette,
France}

{\footnotesize\it $^b$ Laboratoire de Physique Corpusculaire et de Cosmologie,
Coll\`ege de France/ IN2P3 (CNRS), 11, place Marcellin Berthelot, 
F-75231 Paris, France}

\end{flushleft}

\section {Introduction}
\hspace{0.5cm}In a recent presentation at the TAUP97 conference [1]
about the analysis of data from a NaI(Tl) 
100 kg underground detector run at the Gran Sasso Laboratory
by the DAMA/NaI group, 
the speaker concluded there was a hint for a WIMP of mass 60 GeV with
a cross section on proton $\sigma$$_p$ of 10$^{-5}$ pb (Spin Independent
coupling) (see also [2]).\

Even if no claim of any definite signal is made, 
such a statement is strong enough to deserve a critical look.
The present note puts forward a few arguments which
point towards a largely overestimated statistical 
significance of the effect, 
an inconsistency in the energy distribution and several
experimental effects which could easily mimic the observed excess.

\section {About the statistical treatment}
\subsection {The effect is not distributed among the crystals as expected}
\hspace{0.5cm}The main evidence of the "hint" comes from the 
weighted average value, over
the nine crystals constituting the detector, of a variable Sm quantifying
the modulation of the experimental measured rate. This variable Sm, 
weighted over the nine crystals in the 2-12 keV energy interval, 
is found to be 0.034$\pm$0.008 evts/kg/day/keV. This is
considered as a significant signal and interpreted as the
hint for a modulation.\

However, the distribution of the statistical
significance per crystal (weighted excess rate of events divided 
by the statistical error, calculated from the table 2 of ref 2), 
is drastically different from the expected one (figure 1)
in the case of an homogeneously distributed 
effect among the crystals, and it can be seen that
the mean deviation comes from only 3 crystals c7, c8 and c9, while the
distribution for the 6 others is in agreement with no effect.
The $\chi^{2}$ probability that the observed distribution comes from
the expected one is 2 10$^{-4}$.
This raises the important but unadressed question of existing 
systematic effects depending on the crystal.

\ffig{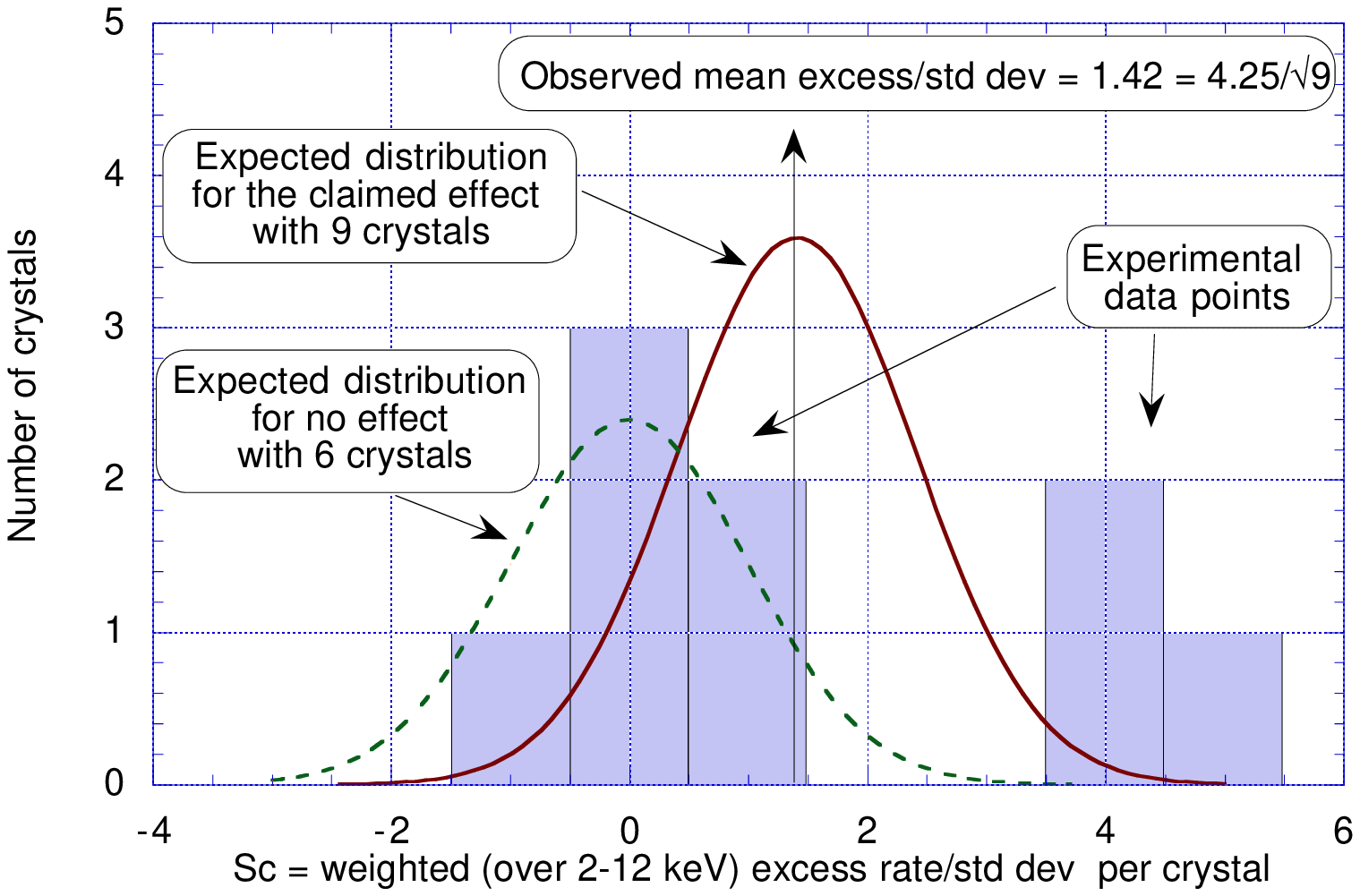}{80mm}{\it 
Experimental (histogram) and expected (full line) Sc crystal distributions, 
where Sc is the statistical significance : 
weighted excess rate/standard deviation for the 2-12 keV region.}{figuremd}

\subsection {No reason to select the particular 2-12 keV region where 
the effect is observed}
\hspace{0.5cm}The energy interval 2-12 keV has been selected because
this is "where the major part of a signal would be expected" [2].
This is too crude a statement. \

It depends on the nature of the
considered WIMP. In the case of Spin Independent interacting WIMP, the
expected interactions on NaI occur mostly on the Iodine nuclei because of the
A$^{2}$ factor ((127/23)$^{2}\simeq$30)  and of the reduced mass factor 
($\mu$$^{2}$ (W,I)/$\mu$$^{2}$(W,Na)$\simeq$6 for a WIMP(W) mass of 60 GeV).
Then the upper value of the energy window  which would keep 90 $\%$
of the modulation of a signal above 2 keV from any
WIMP mass is 6 keV (taking into account the quenching factor, 
the form factor effect and the energy resolution). 
Would it be the Spin Dependent case,
interactions occur mostly on Sodium and this upper value is
around 25 keV.\

So there is no physics grounds in  considering a priori a 2-12 keV region 
which seems to have
been chosen "ad hoc" to enhance the statistical significance of the 
"hint of signal". Anyway, the 3 first data points (2-5 keV region),
divided by the corresponding statistical errors (from the table 3 of ref 2),
are respectively 0.62 $\sigma$, 0.57 $\sigma$ and 1.3 $\sigma$ away
from zero, showing no significant excess, and so no hint of a 60 GeV
WIMP signal. This is illustrated on figure 2
where the experimental Sm energy distribution 
is shown together with the signal from a 60 GeV WIMP, 
the integral of the signal being normalised to the total 
excess in the 2-12 keV region. 
This normalisation, imposed by the data, corresponds to a cross section larger
than the claimed one by more than one order of magnitude,
incompatible with published limits.
If, alternatively, a normalisation to $\sigma$$_p$ = 10$^{-5}$ pb is 
assumed, then more than 90  $\%$ of the observed effect would be unexplained.
So the interpretation of the excess as being due to a 60 GeV WIMP does not
fit the data.

\ffig{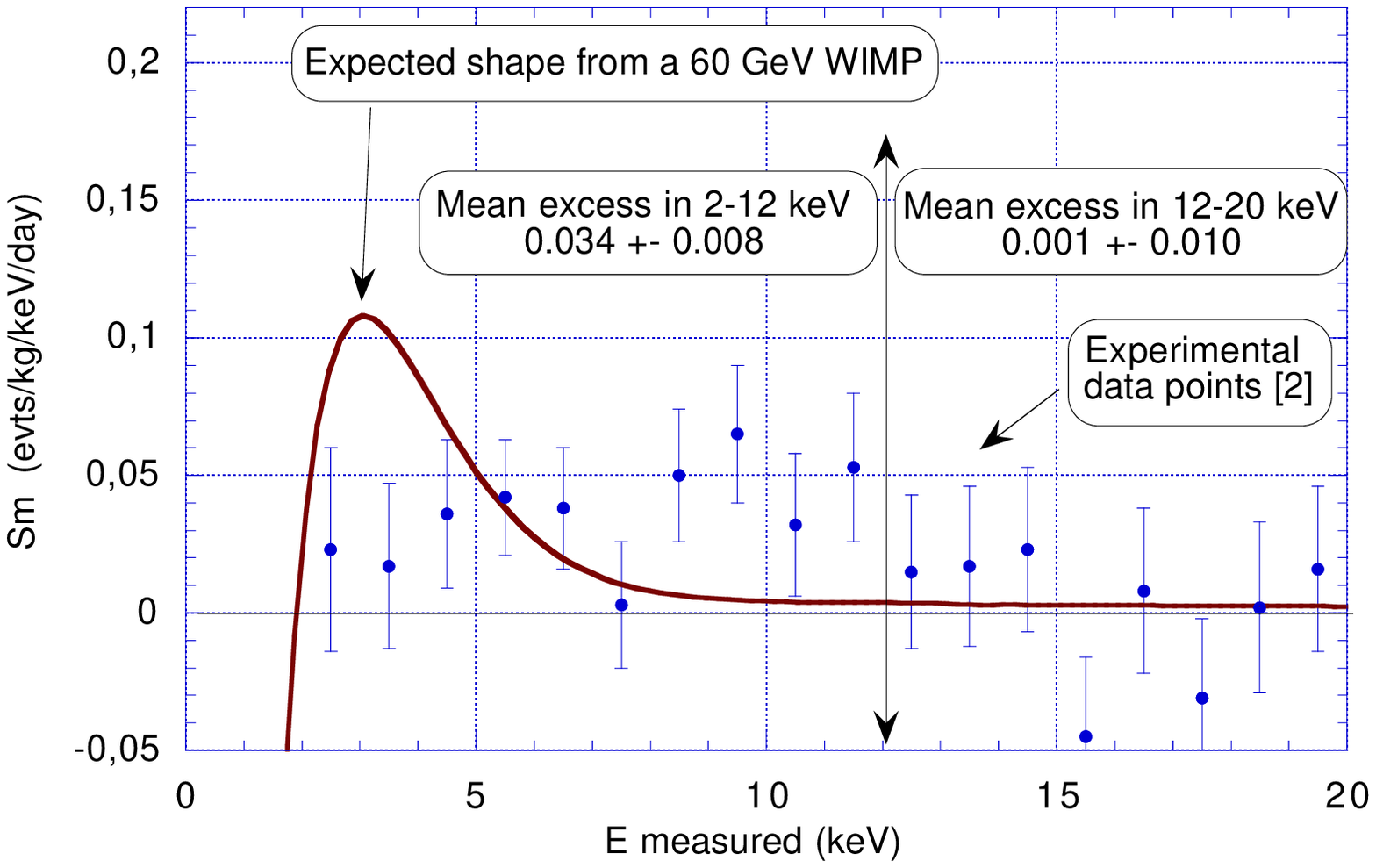}{80mm}{\it Sm energy distributions : experimental
and expected from a 60 GeV WIMP interacting on Iodine, assuming the same
normalisation in the 2-12 keV region.}{figuremd}

\subsection {No discussion on the selection of the time windows}
\hspace{0.5cm}The June time window is very short, only 12 days. On the 
other hand, the quoted spread of the winter time measurements, 
about 70 days, corresponds to twice
the time needed for an exposure of 3368 kg.d with a 115 kg setup. So, either 
there were a shut down of the experiment or data removal. How does
the effect vary with lengths and positions of time windows ? All these points
have direct consequences on the significance of a possible effect.

\section {About the systematic effects}

\subsection {Large subtraction of the PMT noise in the 2-6 keV region 
not under control}
\hspace{0.5cm}With such a set up
(underground conditions, each crystal seen by 2 PMT's) and trigger
(the coincidence of the two PMT's), 
the photomultiplier noise (random coincidences) dominates the counting rate
in the 2-6 keV region and extends up to 10 keV.
As indicated in [3], this noise (with characteristic shape) is 
removed by software cuts. There is however an overlap between the 
noise pulse shape distribution and true NaI(Tl) pulse shape distribution.\

The uncertainty in the removal of this noise, for data taken at
six months time interval, should not exceed the level of 1$\%$ of
the signal (as absolutely needed in any annual modulation
analysis). This implicit hypothesis of stability is here out of control, a
priori not realistic, in any case not discussed.\

Correlatively, the efficiency for applying these cuts, that is the fraction
of true NaI(Tl) events kept, "varies from 30-40 $\%$ 
(depending on the crystal)
up to 100 $\%$ between 2 and 12 keV" [3]. The uncertainty on
these selections and
corrections should also be taken into account.
\subsection{Other systematic effects}
\hspace{0.5cm}There are other systematic effects such as the variation
in time of the energy normalisation, and the decay rate of the
residual contaminations which must be evaluated and the
corresponding uncertainties taken into account
before talking about any possible hint of modulation.

\section{Conclusion}  

\hspace{0.5cm}Using information available from the papers themselves,
it has been shown that in no way the result presented in [1,2] can be
considered as a hint of a WIMP annual modulation.\

\vspace{1cm}
\begin{large}
{\bf References}\\
\end{large}
\noindent  1. P. Belli, talk at TAUP97, Gran Sasso Laboratory,
September 7-11, 1997\\
  2. R. Bernabei et al., preprint ROM2F/97/33 - August 1st,1997 \\
  3. R. Bernabei et al., Phys. Lett. B 389 (1996) 757-766\\

\end{document}